\documentclass[aps,onecolumn,10pt,final, showpacs]{article}
\usepackage{amsmath,amssymb,amsfonts}
\usepackage{graphics,graphicx}
\usepackage[unicode,colorlinks,linkcolor=blue]{hyperref}
\usepackage[cp1251]{inputenc}   
\usepackage{color}
\usepackage{verbatim}
\usepackage{latexsym}
\usepackage{setspace}
\usepackage[english]{babel}
\begin{document}
\title{Cosmological implications of Higgs field fluctuations during inflation}
\author{A.V. Grobov\thanks{alexey.grobov@gmail.com}, R. V. Konoplich \thanks{rk60@nyu.edu}, S.~G.~Rubin\thanks{sergeirubin@list.ru}\\
National Research Nuclear University MEPhI\\
Department of Physics, New York University,\\
Physics Department, Manhattan College}

\date{}

\maketitle

\begin{singlespace}
\small{Cosmological implications of Higgs field fluctuations during inflation are considered.
This study is based on the Standard Model and the standard quadratic model of chaotic inflation
where the Higgs field is minimally coupled to gravity and has no direct coupling to the inflaton.
In the Standard model the
renormalisation group improved effective potential develops an instability (an additional minimum and
maximum) at large field values. It is shown that such a new maximum
should take place at an energy scale above $10^{14} \;\text{GeV},$ otherwise a universe like ours is extremely unlikely.
The extension to the case of the Higgs field interacting with the inflaton field is discussed.}
\end{singlespace}


\section{Introduction}
In 2012, the ATLAS and CMS Collaborations reported the discovery ~\cite{atlas_H, cms_H}
of a new neutral resonance at the LHC. It was demonstrated that the new particle with
mass of around 125 GeV was dominantly produced via the gluon-fusion process and decayed into pairs of gauge
bosons: $\gamma \gamma$, ZZ and WW. Subsequent measurements showed that its dominant spin and parity,
and its couplings to fermions and bosons were compatible within available statistics with expectations for the
Standard Model Higgs boson ~\cite{atlas_coupl, cms_coupl, cms_spin, atlas_spin}.

The discovery of Higgs boson initiated
many studies on the role of the Higgs field in the formation of the early Universe.
In particular, it was found by using the renormalisation group approach that for the measured
values of the Higgs boson and the top quark masses that our Universe (the electroweak vacuum) should be
metastable ~\cite{Espinosa, Degrassi, Strumia}. In the pure SM the instability develops at
energies $\gtrsim 10^{10}$ GeV. Nucleation of
bubbles of true vacuum within the space of our universe should lead to a decay of the electroweak vacuum
to the state of lowest energy but its life-time would exceed the current age of our universe by
many orders of magnitude ~\cite{Strumia}. These results are very sensitive to the Higgs boson mass and especially
to the top quark Yukawa coupling (the top quark mass) ~\cite{Strumia, Bezrusha}. A change of the top quark
mass by a couple of GeV can make the electroweak vacuum stable.

The vacuum instability can become cosmologically relevant at the inflationary
stage ~\cite{Espinosa, Hook, Kobakhidze, Enqvist, Lebedev, Hogan} when
large fluctuations can drag the Higgs field out of the electroweak vacuum to the false vacuum.
For some regions of
parameter space the existence of a universe like ours would be extremely unlikely and new physics would be
required to stabilize the electroweak vacuum.

In this work the cosmological implications of fluctuations of the Higgs field during inflation are considered
and constraints on the location of the false vacuum are derived.

\section{Higgs potential at high energy scale}

\subsection{Instability scale}

In the standard model of chaotic inflation ~\cite{Linde} with the inflaton
scalar field $\phi$ and quadratic potential $V_{inf}(\phi)$ the scale of inflation is
given by the Hubble parameter
\begin{equation}
H \approx 1.1 \cdot 10^{14}\sqrt{r/0.2}  \;\text{GeV}
\end{equation}
where r is the tensor-to-scalar ratio. For models of slow roll inflation ~\cite{Slowroll}.
Planck data gives an upper bound $r<0.11$ at the pivot scale $k=0.002 Mpc^{-1}$ \cite{Planckinfl}.
For Starobinsky inflation ~\cite{Starobinsky} $r \approx 0.003$.
This gives the energy density at inflation
\begin{equation}
V_{inf}^{1/4} \approx \sqrt{M_{pl} H} \approx 3.7 \cdot 10^{16} \cdot (r/0.2)^{1/4}  \sim 10^{16} \;\text{GeV .}
\label{Vinf}
\end{equation}
The maximum of the Higgs effective potential calculated at NLO approximation
~\cite{Schwartz} is
\begin{equation}\label{Vmax}
V_{max}^{1/4} \approx 2.88 \cdot 10^9 \;\text{GeV}.
\end{equation}
Note that the effective potential is gauge-dependent ~\cite{Schwartz, Luzio, Nielsen} as is
the value of the field h. But the value of $V(h)$ at an extremum in $h$ is gauge-invariant.

\begin{figure}[h]
\centering
\includegraphics[width=0.8\textwidth]{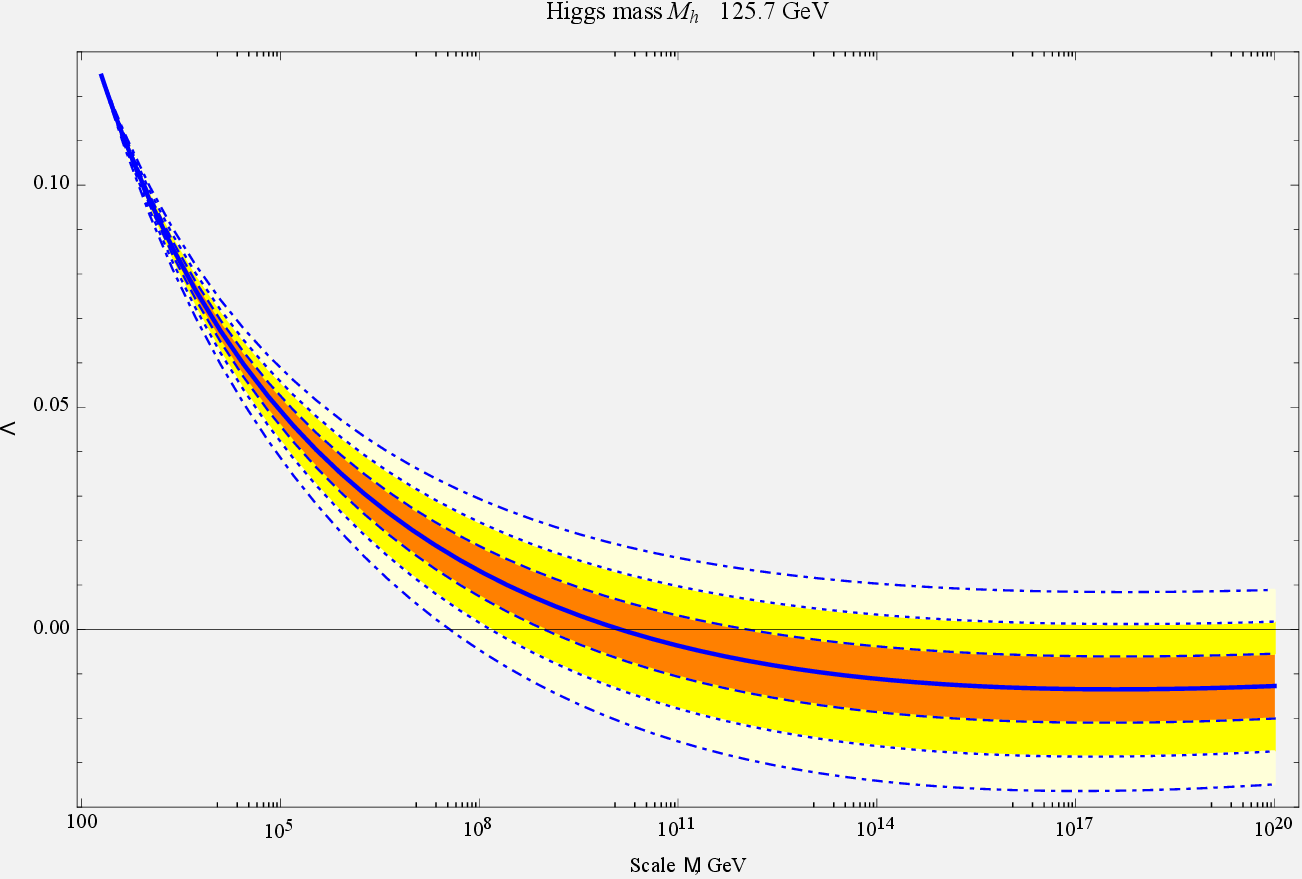}
\caption{Higgs self-coupling $\lambda$, obtained in the framework of $\overline{MS}$
renormalization scheme for central values $M_h = 125.7$ GeV and $m_t = 173.34 \pm 0.82$ GeV.
Deviations for $m_t\pm 1 \sigma$, $m_t\pm 2 \sigma$, $m_t\pm 3 \sigma$ are shown.}
\label{fig:lambda}
\end{figure}

It follows from Eqs.~\ref{Vinf} and \ref{Vmax} that $V_{inf}(\phi) \gg V_{max}(h)$ so that
the presence of the Higgs potential does not affect inflation if the initial Higgs field h is in the range
$0<h<\Lambda_I$ where $\Lambda_I$ is the energy scale at which the Higgs potential becomes unstable.
The instability scale occurs at energies much higher than the electroweak scale. Thus, low power
terms in h in the effective potential can be neglected and the effective potential becomes
\begin{equation}
V(h)=\frac{1}{4}\lambda(h)h^4
\end{equation}
The behavior of the running coupling $\lambda$ in the renormalisation group approach
obtained by using the code ~\cite{Bezrukovcode}
in the renormalisation group approach is shown in Fig.~\ref{fig:lambda}.
The renormalisation group evolution of the running coupling demonstrates that $\lambda$
becomes negative somewhere in the wide energy range of $10^8 \;\text{GeV}$ to $10^{19} \;\text{GeV}$ depending on
the top quark mass (the Yukawa coupling). This is the so called instability scale $\Lambda_I$.
At this scale the potential $V(h)$ is steep so that the position $\Lambda$ of the potential maximum
is of the order of $\Lambda_I$ \cite{Luzio}. The instability scale $\Lambda_I$, defined in $\overline{MS}$
subtraction scheme as the zero point of the running coupling $\lambda$, is
gauge-invariant ~\cite{Bezrusha} and, thus, convenient for estimates.

\subsection{Higgs field evolution after inflation}

The presence of the second minimum in the Higgs effective potential and fluctuations of
the Higgs field h during inflation lead to non-trivial consequences. If $h<\Lambda$, then
at the end of inflation the Higgs field evolves classically to the electroweak vacuum at
$v_1 \approx 246 \;\text{GeV}$. In the opposite case ($h>\Lambda$) the Higgs evolves to the second
minimum $v_2$ at the high energy scale $v_2>>v_1$.
Note that the high temperature thermal effects at the preheating stage could strongly influence this conclusion.

Inflation is the reason for the wide spread of the Higgs field in causally disconnected space regions.
In a time $t \sim H^{-1}$ the field fluctuates by $\delta h \sim H/2\pi$ (1 e-fold).
For the number of e-folds $N_V=60$ the average deviation of the Higgs field from its
initial value during the time of inflation is about $\Delta h_V = \sqrt{N_V}H/2\pi \sim H \simeq 10^{14}\;\text{GeV}$.
In particular,
the Higgs field at the end of inflation in different space
regions can have values either below or above the position of the
second maximum $\Lambda$ of the Higgs potential.
The consequent classical evolution of this field should lead to observable effects related
to the appearance of domain walls ~\cite{Zeldovich, MyBH}.

Let us estimate the time required to stabilize the Higgs field at its observable
minimum $v_1$. Consider the classical equation of motion $\ddot{h}(t)+V'_h (h) = 0$
where the Hubble parameter is supposed to be small after inflation. If the initial Higgs field
value is close to zero, the time of motion of the field to its minimum is estimated as
\begin{equation}
\Delta t \sim \frac{\Delta h}{\sqrt{2V(\Delta h)}} \sim \frac{10^{14}}{10^{19}} \;\text{GeV}^{-1} \sim 10^{-29} \;\text{s}.
\end{equation}
It is assumed here that $V(\Delta h)$ is of the order $V_{max}$.
As a result, the Higgs field quickly reaches its minimum. This occurs much earlier than the nucleosynthesis stage
thus avoiding a contradiction with observations.

The standard model of chaotic inflation leads to the large number $(\sim e^{180})$ ~\cite{Starobinskyinfl1}
of causally disconnected space regions.
Some number $n_1$ of them evolve in a short time interval to the electroweak vacuum $v_1$. The remaining $n_2$
regions evolve to the vacuum $v_2$. Our universe could appear if $n_2<<n_1$.
Small number of regions with the de Sitter space does not pose a threat. For an outside observer they rapidly collapse.

Temperature effects could drastically change this situation. Decay of the inflaton field into SM particles would heat up the plasma.
These temperature effects become relevant when reheating temperature is comparable to the height of potential barrier between two minima
\begin{equation}
T_{reh} \sim V_{max}^{1/4}\approx2.88 \cdot 10^9 \;\text{GeV}.
\end{equation}
An expression for the reheating temperature can be written as \cite{Reheating}
\begin{equation}
T_{reh}=\left(\frac{30\rho_{reh}}{\pi^2 g_{reh}}\right)^{1/4}
\end{equation}
where $\rho_{reh}$ is an energy density of plasma, $g_{reh}$ is the number of degrees of freedom.
$T_{reh}$ could be in the range $10^4$ GeV <$T_{reh}$<$10^{16}$ GeV depending on a specific inflationary model.
For example, in the case of Higgs driven inflation \cite{Bezrusha2} the reheating temperature should exceed $10^{13}$ GeV (depending on the coupling with gravity $\xi H^{+}H R$) to stabilize Higgs potential and erase the second minimum.

\section{Fluctuations of the Higgs field during inflation}

To determine the cosmological implications of inflation for the stability of the
electroweak vacuum consider two different regimes:

1. $H>\Lambda$. It will be shown that in this regime most of the universe will be
in an unacceptable vacuum state at the high energy scale.

2. $H<\Lambda$. In this case, initial Higgs field values are possible for which,
despite the inflationary fluctuations, the Higgs field will reside in the electroweak vacuum
after inflation.

Let us calculate the probability density to find a definite Higgs field value at a given
space point. The Higgs field is a complex two component field. During inflation (quantum regime)
the components of the Higgs field fluctuate randomly. It is known ~\cite{Linde, Starobinsky}
that fluctuations of a scalar field (neglecting its potential) can be described by the
probability density
\begin{equation}
P(\chi , t)=Ae^{-B\cdot (\chi -\chi_0)^2}
\end{equation}
where $B= 2\pi^2 /(H^3 t)$, $A$ is a normalization constant
and $\chi_0$ is the initial value of the field. In the case of the Higgs field this is valid for
all four components $h_i$ of the field. Therefore
\begin{equation}\label{ph2}
P(h_1,h_2,h_3,h_4 , t)=Ae^{-B\sum_i (h_i -h_{0,i})^2}
\end{equation}
where $h_{0,i}$ are initial components of Higgs field.
The probability of finding the Higgs field $h=\sqrt{h^2}=\sqrt{\sum{(h_i)^2}}<\Lambda$
in some space region is given by
\begin{equation}\label{prob}
P_{\Lambda}(t)\equiv \int P(h_1,h_2,h_3,h_4 , t)\theta(\Lambda - h)dh_1dh_2dh_3dh_4 .
\end{equation}

At the end of inflation this probability should be close to one in order for most of the universe
to land in the electroweak vacuum \eqref{ph2}.

Substituting Eq.~(\ref{ph2}) into Eq.~(\ref{prob}) and assuming $h_{0,i}=0,\, i=1,2,3,4$ one obtains

\begin{eqnarray}
P_{\Lambda}=A\int_0 ^{\Lambda} e^{Bh^2}h^3 dh .
\label{plambda}
\end{eqnarray}

The normalization constant $A$ can be found from the equation
\begin{equation}
A\int_0 ^{\Lambda_{b}} e^{-Bh^2}h^3 dh =1
\label{norm}
\end{equation}
with upper limit $\Lambda_b$ to be defined.

Calculation of the normalization constant $A$ is nontrivial because of uncertainty in
definition of the upper limit $\Lambda_b$. It can be obtained by equating a speed
of classical motion and fluctuations during inflation as was done in \cite{Hook}.
Alternatively, it also can be estimated by the virial theorem equating the
kinetic and potential energy of the Higgs field fluctuations \cite{Dolgov92}.
In any case, the shape of the Higgs potential needs to be known to perform estimations.
In spite of this uncertainty, estimates lead to the inequality $\Lambda_{max} <<\
\Lambda_b$ so that the assumption $\Lambda_b \rightarrow \infty$ is reasonable.

We are interested in the probability \eqref{prob} at the time $t_{end}=H^{-1}N_{V}$ with $N_{V}=60$,
when inflation has just finished. A straightforward calculation using Eqs.~\ref{plambda} and \ref{norm}
shows that

\begin{equation}
P_{\Lambda}(t_{end}) = \left( 1-e^{-\frac{2 \pi^2 \Lambda^2}{H^2 N_{V}}}- \frac{2 \pi^2 \Lambda^2}{H^2 N_{V}}
e^{-\frac{2 \pi^2 \Lambda^2}{H^2 N_{V}}}\right) .
\label{prob2}
\end{equation}

A change in a gauge parameter is equivalent to a redefinition of the Higgs field according to the
Nielsen identity \cite{Nielsen}. It can be shown \cite{Espinosa2015} that dealing only with the dominant
gauge dependence of the effective potential the integrated probability is independent of the field rescaling.

\subsection{The case $\Lambda<H$}

\begin{figure}[h]
\centering
\includegraphics[width=0.8\textwidth]{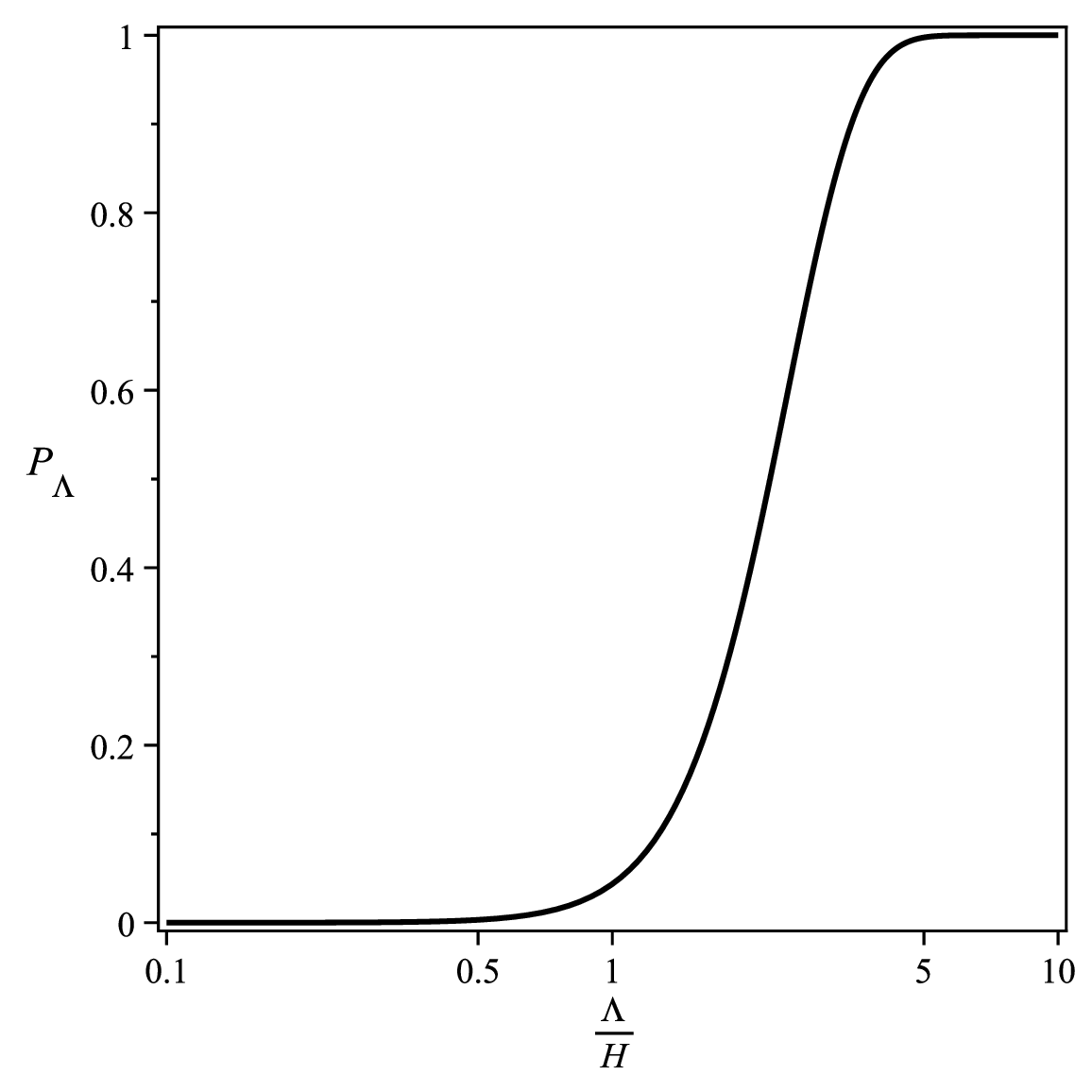}
\caption{Probability \eqref{prob2} of landing in the electroweak vacuum at the end of inflation.}
\label{fig:probability1}
\end{figure}

The probability of landing in the electroweak vacuum at the end of inflation is
shown in Fig.~\ref{fig:probability1}.

It follows from this figure that the probability of finding the Higgs field in the electroweak
vacuum is negligible if $\Lambda<H$. At the end of inflation most space regions
land in the wrong vacuum state $v_2 \gg v_1 \approx 246 \;\text{GeV.}$ In this case regions of the electroweak
vacuum represent bubbles surrounded by a space of different energy density.

If the second minimum of the Higgs potential is negative $(V_{min} < 0)$ then the regions of the vacuum state $v_2$
will form an anti-de Sitter space ~\cite{Hook, Espinosa} in a short time interval. This is because
at the end of inflation the Hubble parameter decreases rapidly and the Higgs field quickly rolls to the vacuum
state $v_2$. In this case the regions of the electroweak vacuum shrink, leading to the dominancy
of anti-de Sitter regions, which in their turn collapse according to classical equations ~\cite{0703146v1}.

If the second minimum is positive $(V_{min} > 0)$ and $\Lambda<H$,
then after the end of inflation the Higgs field is in its vacuum state $v_2$ in most of
space. The probability of quantum tunneling to the
electroweak vacuum in this
case is strongly suppressed because of the width of the potential barrier. As a result our universe -
with the electroweak vacuum - is not realized in this case. Even if $\Lambda$ is close to $H$, the result
is still unacceptable because the coexistence of bubbles of false and true vacuum leads
to the wall dominated universe, in contradiction to observations \cite{Zeldovich}.

Note that the case $V_{min} > 0$ and $\Lambda<H$ seems improbable in the Standard Model because of
the specific behavior of the running coupling $\lambda$. This coupling should change sign from
negative to positive to provide $V_{min} > 0$ and, as can be seen from Fig.~\ref{fig:lambda},
this requires fine tuning of parameters.

Thus, in the scenario $\Lambda<H$ our universe cannot exist.

\subsection{The case $\Lambda>H$}

It follows from Fig.~\ref{fig:probability1} that the existence of our universe requires
$\Lambda > H \sim 10^{14} \;\text{GeV.}$

If $\Lambda\gg H$, the average deviation of the Higgs field from its initial value during the time of inflation $\Delta h_V$ is the order of $H \simeq 10^{14}\;\text{GeV} \ll \Lambda$.
As a result, according to Eq.~(\ref{prob2}), only small number $n_2$ of space regions will be
in the wrong vacuum state $v_2$ if the initial Higgs field $h_{0i}$ were close to zero. However, the
initial Higgs field value at the appearance of the modern horizon is uncertain. If $h_{0i}$
satisfies the condition
\begin{equation}\label{cond3}
|h_{0i}- \Lambda|\leq \Delta h_V
\end{equation}
then field fluctuations could eventually lead to a large number of space regions with
with $h(t_{end})> \Lambda.$ However, if the number of space regions $n_1$ landing in
the electroweak vacuum is still much greater than $n_2$, then it leads to the appearance of our universe.

The presence of regions with vacuum $v_2$ can lead to observable effects such as

1. De Sitter stage continuing in these regions even after the end of inflation.

2. Elementary particle masses being proportional to the vacuum expectation $v_2$.

3. These regions shrink rapidly releasing energy. This could result in local inhomogeneities of the
cosmic microwave background radiation be observed as hot objects with non-standard chemical composition.

\section{Fluctuations of the Higgs field interacting with the inflaton}

In the discussion above the Higgs field had no direct coupling to the inflation
and was minimally coupled to gravity. In this section the discussion is extended to the more general model
including a quartic cross coupling between the Higgs field and the inflaton

\begin{equation}
V=\frac{m^2}{2}\phi^2 +\frac{\lambda(h)}{4}h^4 + g\phi^2 h^2
\end{equation}
where $g\ll 1$. If $m^2\phi^2/2 \gg g\phi^2 h^2$ and
$m^2\phi^2/2 \gg |\lambda(h)|h^4/4$
then the inlaton energy density dominates and inflation is ruled by the inflaton.
The quartic term $g\phi^2 h^2$ in the potential generates during inflation an extra contribution
$m_h^2 =2g\phi^2 \sim 2gM_{Pl}^2$ to the Higgs mass.

The scalar field fluctuations are generated according to \cite{KhloRu, DER} as
\begin{eqnarray}
&&dP(h_{2},t_{1}+\Delta t;h_{1},t_{1}) =dh_{2}\cdot \sqrt{\frac{R}{\pi }
}\exp \left[ -R\left( h_{2}-h_{1}e^{-\mu \Delta t}\right) ^{2}\right] , \nonumber
\label{maindistr} \\ && R = \frac{\mu }{\sigma ^{2}}\frac{1}{1-e^{-2\mu \Delta t}},
\mu =\frac{m_{h}^{2}}{3H},\sigma=\frac{H^{3/2}}{2\pi } .  \nonumber
\end{eqnarray}
where $dP$ is the probability to find the scalar field $h_2$ after a time interval $\Delta t$.

If the initial field value $h_1 =0$ then
after a time interval $\Delta t=H^{-1}$ the variance of  the scalar field is
\begin{equation}
<h^2>=\int h_2 ^2 dP(h_2,H^{-1};0,0)=\frac{1}{2r}\simeq \frac{\sigma^2}{2\mu}=\left[\sqrt{\frac{3}{2}}\frac{H^2}{(2\pi )m_h}\right]^2.
\end{equation}
where it was assumed that $m_h \gtrsim H$.

Inserting $m_h$ into the equation above one finds that
the amplitude of Higgs field fluctuations for one e-fold is given by
$$\delta h =\sqrt{<h^2 >}\sim \frac{H^2}{\sqrt{g}M_{Pl}}\sim \frac{10^9}{\sqrt{g}} GeV$$.
Thus,  if $m_h \gtrsim H$ then the Higgs field fluctuations are suppressed by $M_{Pl}$ .
The opposite case $m_h \lesssim H$ and the case of a non-minimal Higgs coupling to gravity
were considered in \cite{Espinosa2015} and it was shown that
for some regions of model parameters the universe can land in the false vacuum.

\section{Conclusion}
In the Standard Model the location of the electroweak vacuum $v_1 \approx 246 \;\text{GeV}$ is fixed by the Fermi coupling,
which is determined with high precision from muon decay. However, the renormalisation group improved Higgs potential
acquires an additional minimum (and maximum) and develops an instability at large Higgs field values.
This instability is cosmologically relevant.
Depending on the instability scale and initial Higgs field, the fate of the Universe could be drastically different.

In the simple model based on the Standard Model and the standard quadratic model of chaotic inflation
without possible coupling of the Higgs field with the inflaton and/or without the Ricci scalar coupling the situation is as follows.
The field value where the effective potential reaches its maximum $\Lambda$, is close to the instability scale,
due to the steepness of potential. At the inflationary epoch the fluctuations of fields are frozen at the ~$H$ scale.
In the case of $\Lambda<H$ the universe lands in vacuum state which differs from the electroweak one.
The probability of tunneling to the electroweak vacuum is suppressed by the width of the potential barrier.
Calculations performed in this work show that an additional maximum of the Higgs potential
should be located at an energy scale above $10^{14} \;\text{GeV};$ otherwise a universe like ours is extremely unlikely.

Thus, we have shown that the simplest case excludes the existence of our universe in the wide range of parameters and requires
the new physics to step in and protect the electroweak vacuum. However the question of stability of the universe is complicated
due to an interplay of many effects. It looks like more elaborated models including high temperature thermal effects at the reheating stage,
coupling of the Higgs field to the inflaton and/or non-minimal coupling to gravity avoid the danger for a universe like
ours to land in the false vacuum. Various aspects of this problem are discussed in \cite{Bezrusha, Espinosa2015, Bezrusha2, Reheating, Gross, Kear, Kear2}.

The discovery of the Higgs boson and first measurements of its mass have already played an important role in our understanding of the early
universe excluding the simplest types of inflationary models. We will wait for new results on Higgs boson mass and coupling measurements
in Run II and III of the LHC and precision Higgs boson physics measurements from the future high luminosity LHC.

\section{Acknowledgments}
The work of A.G. and S.R. was supported by The Ministry of education and science of Russian Federation, project 3.472.2014/K.
The work of R.K. was partially supported by U.S. National Science Foundation grant No.PHY-1402964.
The authors are grateful to K. Belotsky and F. Bezrukov for fruitful discussions and suggestions.


\begin{thebibliography}{99}

\bibitem{atlas_H}%
G. Aad et al. [ATLAS Collaboration],
Phys. Lett. B \textbf{716}, 1-29 (2012).

\bibitem{cms_H}%
V. Khachatryan et al. [CMS Collaboration],
Phys. Lett. B \textbf{716}, 30-61 (2012).

\bibitem{atlas_coupl}%
G. Aad et al. [ATLAS Collaboration],
Phys. Lett. B \textbf{726}, 88-119 (2013).

\bibitem{cms_coupl}%
V. Khachatryan et al. [CMS Collaboration],
Phys. Rev. Lett. \textbf{110}, 081803 (2013).

  \bibitem{cms_spin}%
V. Khachatryan et al. [CMS Collaboration],
arXiv:1411.3441.

\bibitem{atlas_spin}%
G. Aad et al. [ATLAS Collaboration],
Phys. Lett. B \textbf{726}, 120-144 (2013).

\bibitem{Espinosa}%
J. R. Espinosa,
G. Giudice and
A. Riotto,
JCAP   \textbf{0805}, 002 (2008).

\bibitem{Degrassi}%
G. Degrassi,
S. Di Vita,
J. Elias-Miro,
J. R. Espinosa,
G. F. Giudice,
G. Isidori and
A. Strumia,
JHEP   \textbf{1208}, 098 (2012).

\bibitem{Strumia}%
D. Buttazzo,
G. Degrassi,
P. P. Giardino,
G. F. Giudice,
F. Sala,
A. Salvio, and
A. Strumia,
JHEP   \textbf{1312}, 089 (2013).

  \bibitem{Bezrusha}%
F. Bezrukov and
M. Shaposhnikov,
arXiv:1411.1923.

  \bibitem{Hook}%
A. Hook,
J. Kearney,
B. Shakya, and
K. M. Zurek,
arXiv:1404.5953.

  \bibitem{Kobakhidze}%
A. Kobakhidze and
A. Spencer-Smith,
arXiv:1404.4709.

\bibitem{Enqvist}%
K. Enqvist,
T. Meriniemi and
S. Nurmi,
JCAP   \textbf{1407}, 025 (2014).

\bibitem{Lebedev}%
O. Lebedev and
A. Westphal,
Phys.Lett. B   \textbf{719}, 415-418 (2013).

\bibitem{Hogan}%
M. Fairbairn and
R. Hogan,
Phys. Rev. Lett.   \textbf{112}, 201801 (2014).

\bibitem{Linde}%
A. D. Linde,
Phys. Lett. B   \textbf{129}, 177 (1983).

\bibitem{Slowroll}%
A. D. Linde,
Contemp.Concepts Phys.   \textbf{5}, 1-362 (2005).

  \bibitem{Planckinfl}
Planck collaboration,
arXiv:1502.02114.

  \bibitem{Starobinsky}%
A. A. Starobinsky,
  in:
  Field Theory, Quantum Gravity and Strings
Lecture Notes in Physics, edited by
  H. J.~de Vega and N.~Sanchez,
  246
  (Universite Pierre et Marie Curie, Paris, 1986), pp.\,107-126.

\bibitem{Schwartz}%
A. Andreassen,
W. Frost and
M. D. Schwartz,
Phys. Rev. Lett.   \textbf{113}, 241801 (2014).

\bibitem{Luzio}%
L. Di Luzio and
L. Mihaila,
JHEP   \textbf{1406}, 079 (2014).

\bibitem{Nielsen}%
N. K. Nielsen,
Phys.Rev. D   \textbf{90}, 036008 (2014).

  \bibitem{Bezrukovcode}%
http://www.inr.ac.ru/~fedor/SM/download.php

\bibitem{Zeldovich}%
Ya. B. Zel'dovich,
I. Yu. Kobzarev and
L. B. Okun',
JETP   \textbf{40}, 1 (1975).


\bibitem{MyBH}%
S. G. Rubin,
M. Yu. Khlopov, and
A. S. Sakharov,
Grav.Cosmol.   \textbf{S6}, 51-58 (2000).

\bibitem{Starobinskyinfl1}%
A. A. Starobinsky,
Phys. Lett. B   \textbf{91}, 99-102 (1980).

\bibitem{Dolgov92}%
A. D. Dolgov,
Physics Reports   \textbf{222}, 309-386 (1992).

\bibitem{Espinosa2015}%
J. R. Espinosa,
G. F. Giudice,
E. Morgante,
A. Riotto,
L. Senatore,
A. Strumia and
N. Tetradis,
arXiv:1505.04825.

\bibitem{0703146v1}%
B. Freivogel,
G. T. Horowitz and
S. Shenker,
 JHEP   \textbf{0705}, 090 (2007).

\bibitem{Reheating}%
J. Ellis,
M. Garcia,
D. Nanopoulos and
K. Olive,
arXiv:1505.06986.

\bibitem{Bezrusha2}
F. Bezrukov,
J. Rubio, and
M. Shaposhnikov,
arXiv:1412.3811.

\bibitem{KhloRu}%
M. Yu. Khlopov and
S. G. Rubin,
Cosmological Pattern of Microphysics in the Inflationary Universe
(Kluwer Academic
Publishers, Dordrecht, 2004) vol. 144.

\bibitem{DER}%
V. I. Dokuchaev,
Yu. N. Eroshenko and
S. G. Rubin,
arXiv:0709.0070.

\bibitem{Gross}%
C. Gross,
O. Lebedev, and
M. Zatta,
arXiv:1506.05106.

\bibitem{Kear}%
J. Kearney,H. Yoo and
K. M. Zurek,
arXiv:1503.05193.

\bibitem{Kear2}%
W. E. East, J. Kearney, B. Shakya, H. Yoo, K. M. Zurek,
arXiv:1607.00381

\end{thebibliography}
\end{document}